\documentclass[conference]{IEEEtran}
\IEEEoverridecommandlockouts

\usepackage{cite}
\usepackage{amsmath,amssymb,amsfonts}
\usepackage{algorithmic}
\usepackage{graphicx}
\usepackage{textcomp}
\usepackage{xcolor}
\def\BibTeX{{\rm B\kern-.05em{\sc i\kern-.025em b}\kern-.08emz
    T\kern-.1667em\lower.7ex\hbox{E}\kern-.125emX}}

\usepackage{comment}
\usepackage{lipsum}  
\usepackage{multicol}
\usepackage{multirow}
\usepackage{booktabs}
\usepackage{threeparttable}
\usepackage{circledsteps}
\pgfkeys{/csteps/inner color=black, /csteps/fill color=white}
\usepackage{fontawesome}
\usepackage{subcaption}

\definecolor{deliverable}{HTML}{000000}
\definecolor{note}{HTML}{FF00FF}
\definecolor{remove}{HTML}{FF0000}
\definecolor{new}{HTML}{008000}

\definecolor{note}{HTML}{000000}
\definecolor{new}{HTML}{000000}

\definecolor{dmblue}{HTML}{000000}
\newcommand{\DM}[1]{\textcolor{dmblue}{#1}}

\makeatletter
    \newcommand{\showfont}{
        \begin{itemize}
            \item encoding: \f@encoding{}
            \item family: \f@family{}
            \item series: \f@series{}
            \item shape: \f@shape{}
            \item size: \f@size{}
        \end{itemize}
    }
\makeatother

\begin{document}

% \title{Secure configuration of FPGAs}
% \title{Enabling secure configuration of FPGAs in the Post Quantum Era}
% \title{Enabling Quantum-Resilient Configuration for FPGAs deployed in the Edge}
% \title{Enhancing Secure Configuration of FPGAs with PQC algorithms}
\title{Post-Quantum and Blockchain-Based Attestation for Trusted FPGAs in B5G Networks}
% \title{Secure FPGA Configuration with Post-Quantum Cryptography}

\author{\IEEEauthorblockN{Ilias Papalamprou}
\IEEEauthorblockA{\textit{National Technical University of Athens} \\
\textit{name of organization (of Aff.)}\\
Athens, Greece \\
ipapalambrou@microlab.ntua.gr}
\and
\IEEEauthorblockN{2\textsuperscript{nd} Given Name Surname}
\IEEEauthorblockA{\textit{dept. name of organization (of Aff.)} \\
\textit{name of organization (of Aff.)}\\
City, Country \\
email address or ORCID}
\and
\IEEEauthorblockN{3\textsuperscript{rd} Given Name Surname}
\IEEEauthorblockA{\textit{dept. name of organization (of Aff.)} \\
\textit{name of organization (of Aff.)}\\
City, Country \\
email address or ORCID}
\author{
    \IEEEauthorblockN{
        Ilias Papalamprou\IEEEauthorrefmark{1},
        Nikolaos Fotos\IEEEauthorrefmark{2}, 
        Nikolaos Chatzivasileiadis\IEEEauthorrefmark{2}, 
        Anna Angelogianni\IEEEauthorrefmark{2}, \\
        Dimosthenis Masouros\IEEEauthorrefmark{1},
        Dimitrios Soudris\IEEEauthorrefmark{1}
    }
    \IEEEauthorblockA{
        \IEEEauthorrefmark{1}National Technical University of Athens, Greece
    }
    \IEEEauthorblockA{
        \IEEEauthorrefmark{2}Ubitech Ltd., Digital Security \& Trusted Computing Group, Greece
    }
    \IEEEauthorblockA{
        \IEEEauthorrefmark{1}\{ipapalambrou, dmasouros, dsoudris\}@microlab.ntua.gr,
        \IEEEauthorrefmark{2}\{nfotos, nchatzivasileiadis, angelogianni\}@ubitech.eu  
        }
}

}

\maketitle

\begin{abstract}
The advent of 5G and beyond has brought increased performance networks, facilitating the deployment of services closer to the user. To meet performance requirements such services require specialized hardware, such as Field Programmable Gate Arrays (FPGAs).
% Field Programmable Gate Arrays (FPGAs) are becoming an increasingly popular platform\DM{XXX-na mpei kai kati gia 5G-XXX} for edge computing in recent years. 
However, FPGAs are often deployed in unprotected environments, leaving the user's applications vulnerable to multiple attacks. 
With the rise of quantum computing, which threatens the integrity of widely-used cryptographic algorithms, the need for a robust security infrastructure is even more crucial.
In this paper we introduce a hybrid hardware-software solution utilizing remote attestation to securely configure FPGAs, while integrating Post-Quantum Cryptographic (PQC) algorithms for enhanced security. Additionally, to enable trustworthiness across the whole edge computing continuum, our solution integrates a blockchain infrastructure, ensuring the secure storage of any security evidence.
%Our evaluation demonstrates that some combinations of PQC algorithms, can even reduce the attestation time by \textcolor{note}{XXX} compared to traditional cryptographic methods. 
We evaluate the proposed secure configuration process under different PQC algorithms in two FPGA families, showcasing only $2\%$ overheard compared to the non PQC approach.
% \DM{XXX-edw prepei na mpei kana noumero apo eval-XXX}
\end{abstract}


\section{Introduction}

The rise of 5G and beyond 5G (B5G) networks promises unprecedented improvements in network performance, offering ultra-low latency~\cite{popovski2019wireless}, higher bandwidth~\cite{abedin2018resource}, and support for real-time applications such as autonomous vehicles, augmented reality (AR), and others~\cite{samdanis2020road}.
However, traditional network infrastructures, which rely on specialized hardware optimized for specific tasks (e.g., data packet processing and routing), struggle to meet these demands due to their lack of flexibility, scalability, and manual management complexity~\cite{benson2009unraveling}. 
As a result, network infrastructures are transitioning from specific-purpose hardware to general-purpose processing nodes that leverage virtualization technologies and software-defined networking (SDN)~\cite{kreutz2014software}. 
This new infrastructure not only hosts network-related services but also supports user-developed applications to be deployed closer to the \textit{network edge}, thus, enabling real-time processing required to meet the performance requirements of modern applications~\cite{hu2015mobile}.

In this landscape, hardware accelerators such as GPUs~\cite{abbasi2020efficient} and FPGAs~\cite{chamola2020fpga}, DPUs~\cite{barsellotti2022introducing} and others are being adopted to ensure that performance and real-time requirements are met for latency-sensitive applications deployed at the network edge.
Among these, FPGAs are particularly notable for their reconfigurability and energy efficiency, making them ideal for dynamic edge environments where workloads can vary significantly. 
However, despite their performance advantages, the incorporation of accelerators along with the distributed nature of B5G architectures introduces a range of security challenges that must be addressed to safeguard their deployment.

One of the main security challenges lies in the vulnerability of FPGAs, which are often deployed in environments without adequate protection, leaving them exposed to threats such as malware injections~\cite{la2021denial}, hardware trojans~\cite{chakraborty2013hardware} and others.
Even existing security features, such as AMD's bitstream encryption have been found to be unreliable~\cite{ender2020unpatchable}. 
The challenge of securing FPGAs becomes even more critical with the advent of \textit{quantum computing}, which introduces new security threats. 
If quantum computers achieve their anticipated performance, they could potentially break widely used cryptographic algorithms such as the Rivest-Shamir-Adleman (RSA) system and Elliptic Curve Cryptography (ECC) schemes~\cite{bernstein2017post}, both of which underpin many secure configuration mechanisms today.

Beyond hardware vulnerabilities, the distributed and multi-entity nature of 5G/B5G networks introduces additional security challenges related to trust and authentication. 
In these networks, multiple parties -- such as \textit{infrastructure providers}, \textit{application developers}, and \textit{network operators} -- are involved in the operation of edge computing environments. 
Each entity plays a distinct role, creating a complex ecosystem where verifying the integrity and authenticity of every participant is crucial to ensuring overall network security. 
To address these challenges, robust attestation is needed to verify the actions and trustworthiness of each party, along with immutable storage systems to securely gather evidence.

Although previous research has developed various methods to enhance secure configuration of FPGAs, these approaches do not offer countermeasures against quantum attacks or provide a reliable storage system for collected security evidence. 
For instance, researchers have concentrated on FPGA Trusted Execution Environments (TEEs)~\cite{zhao2022shef,oh2021meetgo,wang2024towards} to ensure secure configuration, with some introducing additional functionalities such as a netlist scanner~\cite{zeitouni2021trusted}, for detecting malicious modules. 
However, all of these solutions rely on ECC-based algorithms, leaving them vulnerable to quantum attacks.
Additionally, all of the previously mentioned solutions overlook the collection and management of attestation evidence, thereby lacking a trustworthy data consistency mechanism. 
While some recent efforts have begun integrating such mechanisms into generic edge environments~\cite{zhang2024bcae}, they have yet to focus on FPGAs.

\DM{In this paper, we propose a remote attestation protocol that integrates Post-Quantum Cryptography (PQC) with blockchain technology to establish a quantum-resistant, trusted attestation framework for FPGAs in B5G networks.
Our approach addresses both the quantum security vulnerabilities of traditional cryptographic methods and the need for reliable, immutable storage of security evidence in multi-entity environments.
The key contributions of this work are:
\textit{i)} We introduce a hybrid hardware and software solution based on remote attestation with PQC algorithms, ensuring that the attestation process remains secure against future quantum threats; 
\textit{ii)} We integrate a blockchain infrastructure, for collecting the security evidence from the deployed FPGA-based edge nodes, providing a decentralized, tamper-resistant ledger for storing attestation records; and \textit{iii)} We evaluate the performance in terms of execution time of our system, based on different PQC algorithms on two FPGA families. Our results shows that for a particular selection of PQC algorithms, minimal overhead is added ($\sim 2\%$) compared to the regular non-PQC approach.}
\section{Proposed Remote Attestation Scheme}
\label{sec:implementation}

\DM{Our solution leverages a custom \emph{remote attestation} scheme, enhanced with PQC to ensure quantum-resilient integrity verification and Blockchain technology to provide an immutable and transparent record of attestation evidence.
Remote attestation enables a system and its components to prove their trustworthiness to a remote \textit{verifier}.}
% \textcolor{note}{After an offline preparation, }the process starts with the \textit{verifier} sending a challenge to the system, which then computes security evidence and sends them to the \textit{verifier} for validation~\cite{coker2011principles}. 
\textcolor{note}{After an offline preparation, }the process starts with the \textit{verifier} sending a challenge to the system, which then sends evidence back to the \textit{verifier} for validation~\cite{coker2011principles}. 
Afterwards, all attestation evidence is stored in a secure storage mechanism i.e., the Blockchain.
% \DM{edw les oti to process ksekinaei me challenge apo ton \textit{verifier} alla sto protocol den einai etsi h seira. -- recheck}

% Furthermore, we enhance the security of the system by including a decentralized storage mechanism, for collecting any remote attestation evidence. 
% Furthermore, by integrating PQC algorithms, we enhance the system's security, while a decentralized storage i.e., the blockchain, collects any remote attestation evidence.
% In the following sections we analyze the overall architecture of or parties involved in our system.

\subsection{System's Architecture}
\begin{figure}[t]
    \centering
    \includegraphics[width=1.0\linewidth]{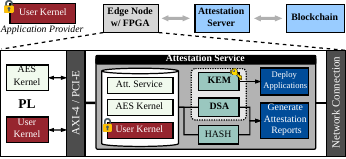}
    \caption{Overview of our proposed approach for secure configuration of edge nodes with FPGAs.}
    \label{fig:arch}
    \vspace{-4ex}
\end{figure}

Figure~\ref{fig:arch} shows the architecture of our system, that consists of three entities: i) the application provider, ii), the infrastructure provider, and iii), the service provider.
\subsubsection{Application Provider}
\DM{The application provider refers to the entity offering the application (bitstream) to be deployed on the FPGA.
This could be either an application related to network functionality (e.g., packet processing) or an end-user application requiring real-time processing (e.g., autonomous driving).
The objective is to ensure that the bitstream executing on the FPGA is an unaltered and authenticated version of the original bitstream provided by the application provider, with no modifications introduced by unauthorized entities.} % \textcolor{note}{XXX-auto den paei infra provider?-XXX}
% In 5G/B5G networks, a wide range of services is deployed at edge nodes, supporting both single-tenant and multi-tenant configurations. These edge nodes, equipped with FPGAs, run applications developed and provided by specialized application vendors.
\subsubsection{Infrastructure Provider}
The infrastructure provider is responsible for delivering and maintaining the underlying hardware and software resources necessary to support the edge computing environments. \textcolor{new}{Unlike other entities, it focuses specifically on the physical and virtual infrastructure to ensure efficient, secure and reliable service operations.}
% \DM{XXX-edw mia protash pws/giati mporei na diaferei auto to entity apo ta alla 2 kai poio einai to objective gia auto to entity-XXX}.
\newline\textbf{Edge Node:} The edge computing node equipped with the FPGA, with its architecture shown in Fig.~\ref{fig:arch}.
It contains hardware (i.e., AES Kernel) and software components (i.e., attestation service) supplied by the infrastructure provider.
The \emph{attestation service} is the core of the edge node and is responsible for managing all tasks for the secure programming of the FPGA; including connection with the attestation server, gathering the security evidence for the remote attestation (Digital Signature Algorithm (DSA), hash functions, communication with AES Kernel), as well as loading the target application onto the FPGA (Key Encapsulation Mechanism (KEM), bitstream decryption).
In contrast to prior works, we focus on enhancing the system's security by utilizing PQC algorithms. 
Specifically, for the KEM and the DSA we avoid regular ECC-based approaches. Their security relies on the difficulty of solving particular mathematical problems (e.g., Discrete Logarithm Problem), which can be solved by quantum computers.
In contrast, PQC algorithms rely on different mathematical properties (e.g., lattices), which remain resistant even to quantum attacks~\cite{chen2016report}.
Furthermore, given the diversity of devices in edge computing environments, our solution is versatile, targeting two distinct FPGAs: (i) PCI-Express (\mbox{PCI-E}) FPGAs with an x86 CPU as the Processing System (PS), which are better suited for larger-scale edge nodes, (ii) System-on-Chip (SoC) FPGAs, which integrate both the Programmable Logic (PL) and the PS within a single system. 
We note that for the PS/PL communication the \mbox{AXI-4} protocol is used.
These are ideal for far-edge environments with tighter energy constraints.
% We note, that these two FPGA families, feature different PS/PL communication: 
% In case of the FPGA accelerator card the PCI-E protocol is utilized, while in SoC FPGAs the \mbox{AXI-4} protocol is used.
\subsubsection{Service Provider}
% Refers to the stakeholder responsible for managing the security and validation services within the 5G/B5G network infrastructure.
\textcolor{new}{This entity is responsible for managing the security and validation services within 5G/B5G network infrastructure, particularly in the role of the \emph{verifier} in the attestation process. In this context, Mobile Network Operators (MNOs) commonly fulfill this role.}
% \DM{XXX-ara einai o \textit{verifier?}-XXX. epishs 1 protash poios mporei na einai auto to entity}
\newline\textbf{Attestation Server:} An external server acting as the Verifier in the remote attestation. Is responsible for validating the values received by the edge node, with pre-stored reference values, that have been acquired from the application and infrastructure provider. Lastly, it is connected with the blockchain.
\newline\textbf{Blockchain:} Consitues the decentralized storage for storing any evidence collected from the attestation requests. 
% In order to align with the ongoing trend towards Self Sovereign Identity, 
The blockchain architecture is based on Hyperledger BESU~\cite{besu}, enabling the design of an Ethereum-based permissioned ledger. 
% \begin{figure}[t]
%     \centering
%     \includegraphics[width=1.0\linewidth]{figures/arch.pdf}
%     \caption{Overview of our proposed approach for secure configuration of edge nodes with FPGAs.}
%     \label{fig:arch}
%     \vspace{-3ex}
% \end{figure}
% \textcolor{remove}{In addition, the use of W3C Verifiable Credentials (VCs) is enriched to record trust-related information on the blockchain in an interoperable and verifiable manner.}
\subsubsection{Trust model}
We assume the presence of an secure attestation server, that manages the secure storage and acquisition of reference values.
% , as well as handling the communication with the edge node and the blockchain infrastructure.
We do not blindly trust the edge node provider; we verify the infrastructure prior to each user's application deployment. 
Lastly, although we acknowledge the risk of physical attacks within the FPGA (e.g., side-channel attacks), they are beyond the scope of this work.

% \textcolor{blue}{Our method is based upon remote attestation, in which the Verifier (in our case the Attestation Server) initiates the process by sending a challenge to the target device, i.e. the FPGA. Multiple values are being calculated and transferred back as a response, in order to get verified.} 
% In the (far) edge node, both software and hardware components are used as shown in Fig.~\ref{fig:arch}, that are responsible for executing the various tasks of the remote attestation, including hashing and encryption/decryption algorithms, key derivation functions etc.

% In our system, we assume the presence of a secure attestation server, that manages the secure storage of reference values, as well as handling the communication with the (far) edge node and the blockchain infrastructure.
% Furthermore, we do not blindly trust the FPGA provider; as we will analyze later in this Section, we verify the infrastructure prior to each user's application deployment. 
% Lastly, although we acknowledge the risk of physical attacks within the FPGA (such as side-channel attacks), these are beyond the scope of this study.

\subsection{Remote Attestation Protocol}

\begin{figure}
    \centering
    \includegraphics[width=1.0\linewidth]{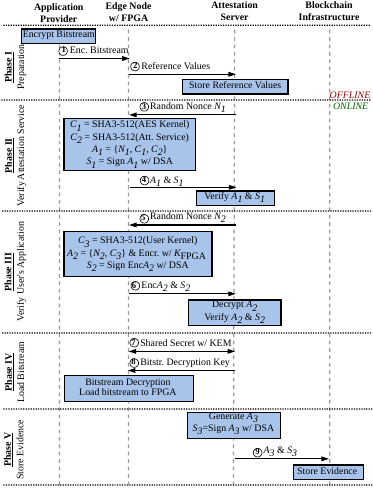}
    \caption{Proposed remote attestation protocol.}
    \label{fig:protocol}
    \vspace{-3ex}
\end{figure}

Fig.~\ref{fig:protocol} shows in detail each step of the proposed remote attestation protocol.
The entire process for securely deploying applications in the edge can be broken down into five phases:
% \par\textit{\underline{Phase I:}} Initially, an offline preparation is required: In step \Circled{1} the user encrypts the application's bitstream with key $K_{btstr}$ and transfers them to the attestation server.
\par\textit{\underline{Phase I:}} Initially, an offline preparation is required: In step \Circled{1} the application provider using $K_{btstr}$ encrypts the application's bitstream and transfers it to the edge node.
% In step \Circled{2} the attestation server acquires from the infrastructure provider the reference values for the deployed application and the attestation modules in the edge node. 
In step \Circled{2} the attestation server acquires from the infrastructure provider all the reference values required for attestation. 
% We also note that a secure network connection is established between the Attestation Server and the (far) edge node, based on Secure Socket Layer (SSL).
Additionally, the infrastructure provider is responsible for generating key pair $\{K_{s,pub}, K_{s,priv}\}$ for the DSA, as well as sharing the public key $K_{s,pub}$ with the attestation server. Lastly, the attestation server collects from the infrastructure provider $K_{FPGA}$ key for the corresponding edge node.  
\par\textit{\underline{Phase II:}} To deploy an application to the edge, the application provider sends a request to the infrastructure provider for initiating the verification process. 
First, we ensure the integrity of the attestation service of the edge node. 
In step \Circled{3} the attestation server generates a random nonce $N_1$ and sends it to the edge node. 
There, the attestation report $A_1$ is produced, containing two SHA3-512 checksum values: for the software ($C_1$) and for the hardware components ($C_2$) of the attestation service. 
The received nonce $N_1$ is appended to $A_1$, producing ${A_1 = \{N_1 \| C_1 \| C_2\}}$ and signed using $K_{s,priv}$, generating $S_1$.
In step \Circled{4} $A_1$ and $S_1$ are transferred to the attestation server, which verifies each received data. 
Upon successful validation, all the components of the attestation service are verified and the remote attestation continues.
\par\textit{\underline{Phase III:}} Following a similar process as in phase II, we proceed with verifying the application's provider bitstream. In step \Circled{5}, the attestation server generates a fresh nonce $N_2$ and sends it to the edge node, where a new attestation report $A_2$ is generated, containing $N_2$ and the encrypted bitstream SHA3-512 checksum $C_3$. Afterwards, the attestation report is encrypted with AES256-CBC in the FPGA, using a pre-installed key $K_{FPGA}$ by the provider, producing ${Enc{A_2} = Enc_{K_{FPGA}}\{N_i \| C_3\}}$. It is then signed with $K_{s,priv}$, producing $S_2$ and in step \Circled{6}, $Enc{A_2}$ and $S_2$ are transferred to the attestation server. 
% Note that, for higher security assurance, a different AES core mode can be used, such as AES256-CBC.
Upon receiving all data, the attestation server decrypts $Enc{A_2}$ using the pre-stored $K_{FPGA}$ associated with the respective edge node. If the verification of both the checksum and the received signature is successful, the protocol proceeds with loading the user's bitstream onto the FPGA. 
% Note that in order to enhance the security of the system, a Physical Unclonable Function (PUF) can be integrated in the FPGA (e.g., from~\cite{streit2021choice}), for securely generating keys.
Note that for higher security assurance, a Physical Unclonable Function (PUF) can be integrated into the FPGA (e.g., from~\cite{streit2021choice}), for securely generating keys.
% In such case, the attestation server instead of obtaining the $K_{FPGA}$, would have to acquire and store multiple PUF responses.
\par\textit{\underline{Phase IV:}} With successful attestation of all individual components, the attestation server sends to the edge node the bitstream decryption key $K_{btstr}$. For securely transferring $K_{btstr}$, a Key Encapsulation Method (KEM) is used for generating the shared secret $K_{ss}$ that acts as the key for the AES256-CBC encryption/decryption of $K_{btstr}$. The edge node consequently receives the $Enc_{K_{ss}}(K_{btstr})$, decrypts the user's bitstream with $K_{btstr}$ and programs the FPGA. 
\par\textit{\underline{Phase V:}} After each remote attestation is completed, whether successful or not, the attestation server forwards the results to the Blockchain.
Specifically, the attestation server generates a new report $A_3$, that includes a fresh nonce $N_3$, as well as the attestation results for both the attestation service (phase II) and the application's provider bitstream (phase III). The report is signed producing $S_3$ and then in step \Circled{9} both $A_3$ and $S_3$ are forwarded to the Blockchain. 
\section{Evaluation \& Discussion}

% This section presents the performance evaluation of securely configuring FPGAs when using different PQC algorithms. Additionally, we discuss how our proposed approach addresses and mitigates potential security threats that could impact the system.

\subsection{Experimental Setup}

% In Table~\ref{tab:specs} the specifications of the two edge nodes are listed. 
In Table~\ref{tab:specs} the specifications of our experimental setup are listed. As outlined in Section~\ref{sec:implementation}, two FPGA families are evaluated; (a) PCI-E FPGA for larger scale edge nodes, and (b) SoC FPGAs for far-edge nodes. 
The software components are a combination of Python 3 and C++, while the hardware modules utilize both High Level Synthesis (HLS) and Register-Transfer Level (RTL) components and are implemented using Vivado/Vitis 2021.1. 
% \textcolor{note}{Additionally, for the software-based cryptographic primitives we leverage the Python libraries \texttt{pypqc} and \texttt{fastecdsa}.}
% Additionally, we utilized Xilinx Runtime (XRT) to manage the PS/PL communication.
% The AES kernel in the FPGA, utilizes only a small percentage of the available FPGA resources (\% across all key resources for ALVEO U280, \% across all key resources for MPSoC ZCU104), while operating at \textcolor{note}{XXXXX} and \textcolor{note}{XXXX} MHz in the ALVEO U280 and MPSoC ZCU104 respectively.
As a proof of concept we utilized an AI analytics accelerator developed in HLS for the deployed application. % The bitstream size is approximately \textcolor{note}{XXX} MBytes. 
The external attestation server is a general purpose x86-based PC, while the Blockchain is deployed on an Ubuntu 22.04 Virtual Machine (VM).

\begin{table}
    \caption{Edge Node \& Far Edge Node Specifications}
    \centering
    \resizebox{1.0\linewidth}{!}{
    \begin{tabular}{cccc}
        \toprule
        \textbf{Specification}                  & \textbf{Edge Node}                & \textbf{Far-Edge Node}        \\ \midrule
        % FPGA                                    & PCI-Express FPGA                  & ZCU104 MPSoC                  \\ 
        % PL                                      & ALVEO U280                        & XCZU7EV                       \\
        FPGA                                    & PCI-E ALVEO U280                  & MPSoC ZCU104 (XCZU7EV)        \\
        \multirow{2}{*}{Utilization$^\dag$}     & LUT=$9\%$, FF=$6\%$               & LUT=$15\%$, FF=$12\%$         \\
                                                & DSP$<1\%$, BRAM=$12\%$            & DSP=$0\%$, BRAM=$17\%$        \\
        \multirow{2}{*}{CPU}                    & Intel Xeon Gold 6530 @ 2.1GHz     & ARM Cortex-A53 @ 1.2GHz       \\
                                                & (x86) \& 256GB RAM                & (aarch64) \& 2GB RAM          \\
        Host OS                                 & Ubuntu Server 22.04 (x86)         & Petalinux 2021.1 (aarch64)    \\
        \toprule
    \end{tabular}
    }
    \begin{tablenotes} \footnotesize
        \item $^\dag$ Includes the AES kernel and the communication between PS/PL.
    \end{tablenotes}
    \label{tab:specs}
\end{table}

% We performed an exploration, regarding the execution time of using different PQC algorithms. In Table~\ref{tab:algorithms} the configurations are shown.
% We performed an exploration regarding the execution time of using PQC algorithms for the DSA and KEM steps, with the different configurations shown in Table~\ref{tab:algorithms}.

\subsection{Performance Evaluation}

\begin{table}
    \caption{Cryptographic Algorithms for each Configuration}
    \centering
    \resizebox{1.0\linewidth}{!}{
    \begin{tabular}{ccccc}
        \toprule
        \multirow{2}{*}{\textbf{Configuration}} & \multicolumn{2}{c}{\textbf{DSA}} & \multicolumn{2}{c}{\textbf{KEM}} \\
        \cmidrule(lr){2-3} \cmidrule(lr){4-5}
                                                & \textbf{Algorithm} & \textbf{Security}$^\dag$  & \textbf{Algorithm} & \textbf{Security} \\ \midrule
        No-PQ                                   & ECDSA              & \faCircleO    & ECDH            & \faCircleO  \\ 
        PQ-I                                    & Falcon 1024        & \faAdjust     & Kyber 1024      & \faCircle   \\ 
        PQ-II                                   & Falcon 1024        & \faAdjust     & McEliece-348864 & \faAdjust   \\ 
        PQ-III                                  & Dilithium 5        & \faCircle     & Kyber 1024      & \faCircle   \\ 
        PQ-IV                                   & Dilithium 5        & \faCircle     & McEliece-348864 & \faAdjust   \\ 
        \toprule
    \end{tabular}
    }
    \begin{tablenotes} \footnotesize 
        \item $^\dag$ \faCircleO \;Low Security Level, \faAdjust \;Mid Security Level, \faCircle \;High Security Level. The security levels against quantum attacks are based on~\cite{nejatollahi2019post,malina2023deploying}.
    \end{tablenotes}
    \label{tab:algorithms}
    \vspace{-4ex}
\end{table}

\begin{figure}
    \centering
    \begin{subfigure}{0.70\linewidth}
        \centering
        \includegraphics[width=1.00\linewidth]{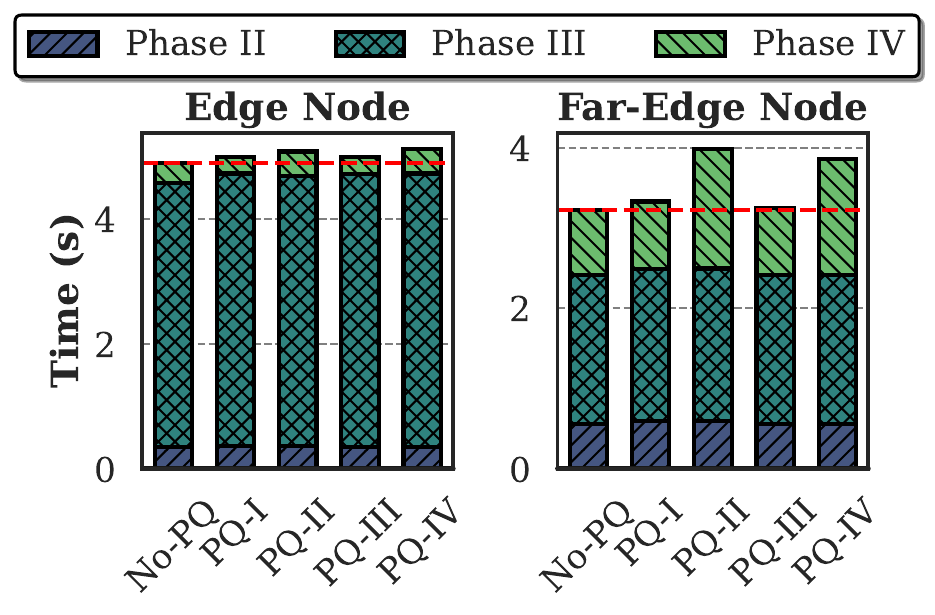}
        \caption{Remote attestation protocol}
        \vspace{-1ex}
        \label{fig:exec_time:attestation}
    \end{subfigure}
    \begin{subfigure}{0.25\linewidth}
        \centering
        \includegraphics[width=1.00\linewidth]{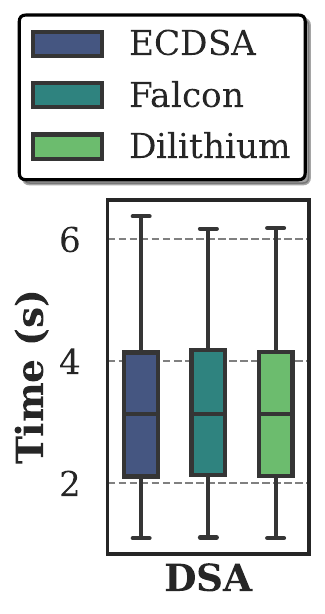}
        \caption{Blockchain}
        \vspace{-1ex}
        \label{fig:exec_time:blockchain}
    \end{subfigure}
    \caption{Execution time for 100 requests.}
    \label{fig:exec_time}
    \vspace{-3ex}
\end{figure}

The proposed solution is evaluated over five configurations based on the cryptographic algorithms selected, as shown in Table~\ref{tab:algorithms}.
We analyze the execution time of (a) the individual phases of the remote attestation (Fig.~\ref{fig:exec_time:attestation}) and (b) pushing data to the blockchain (Fig.~\ref{fig:exec_time:blockchain}).
As a baseline for comparing the PQC algorithms, we choose widely used ECC algorithms. %Our observations are the following: 
% \textcolor{note}{MHPWS NA GRAFTOUN KAI GIA TA DYO FPGA MAZI????}
% We evaluated the execution time of remote attestation and the process of pushing data to the blockchain under different DSA and KEM, as shown in Table~\ref{tab:algorithms}. Our baseline for comparison acts a typical configuration based on ECC that is not PQ resistant. 
% Our observations are the following:

% From Fig.~\ref{fig:exec_time:attestation} we observe that the choosing different KEM has more impact in the execution time than changing the DSA. 
\par\noindent\emph{Edge Node:} Both \mbox{PQ-I\&III} configurations perform similarly with the baseline, with \mbox{PQ-III} showing the lowest overhead at $\sim 2\%$ over the No-PQ configuration. When analyzing each attestation phase individually, phase II shows similar performance across all configurations, regardless of the DSA. Phase III accounts for the majority of the execution time, due to the lengthy process of configuring the Alveo with the AES kernel, which takes  $\sim 3.8$sec. In phase IV, where different KEMs are employed, Kyber outperforms McEliece.
\begin{comment}
Among the different configurations, both \mbox{PQ-I\&III} perform faster than the baseline, with \mbox{PQ-III} achieving the best performance ($XXXX\%$ lower than the baseline).
Furthermore, regarding DSA (Phase II\&III) small variations in execution time are noticed. For KEM (Phase IV), Kyber outperforms all methods, even the classic ECDSA, while a small overhead is observed when utilizing Mceliece method.
\textcolor{note}{We note that the execution time of Phase III is larger the other phases. This is due to the time required for configuring the FPGA with the AES core, that takes $\sim 3.8$ sec.}
\end{comment}

\par\noindent\emph{Far-Edge Node:} 
% \textcolor{red}{Similar results are observed with the smaller edge node.} 
The \mbox{PQ-I\&III} configurations have the closest performance to the baseline, with \mbox{PQ-III} obtaining the best results. 
All configurations in phases II\&III have similar execution times, while in phase IV significant overhead is noticed when using the McEliece method. As with the larger edge node, Kyber emerges as the fastest PQ KEM, while for DSA, both Falcon and Dilithium deliver similar performance.
\begin{comment}
Both \mbox{PQ-I\&III} have similar execution time with the \mbox{Non-PQC} configuration, with the \mbox{PQ-III} performing the best. Additionally, as shown in the case of the edge node, the Mceliece KEM (\mbox{PQ-II\&IV}) increases significantly the execution time.
\end{comment}

\par\noindent\emph{Blockchain:} % From the box plot in Fig.~\ref{fig:exec_time:blockchain}, for all DSA algorithms the time required for uploading the results is almost identical. Furthermore, we observe that the time of each data push varies, since the time required for finalizing each block processing in the blockchain is not constant. Results shown are for 100 runs.
In Fig.~\ref{fig:exec_time:blockchain}, the time required to upload the results is consistent across all DSAs. Additionally, we observe some variation in each data push time, as the time needed to finalize each block's processing in the blockchain is not constant.

% Overall, from the experimental results and considering the security levels shown in Table~\ref{tab:specs}, we conclude that the most prominent configuration is the PQ-III, that features Dilithium and Kyber as the DSA and KEM respectively and has less than $<\%$ overhead over the traditional non PQC solution.
Overall, based on the experimental results and the security levels in Table~\ref{tab:specs}, we conclude that the most prominent configuration is PQ-III, that uses Dilithium as DSA and Kyber as KEM. This selection incurs approximately $2\%$ overhead compared to the non-PQC solution in both FPGA families.

% From Fig.~\ref{fig:exec_time:attestation} we observe that in the edge FPGA (left) the overheads of the PQ algorithms are minimal, with two configurations (PQ-I \& PQ-III) performing faster than our baseline. This is caused by Phase IV, that includes the KEM. In those configurations Kyber is selected, that given it's lightweight design, it outperforms the ECDH algorithm. Furthermore regarding McEliece KEM, it has the largest overhead over the other methods, especially in the case of the Far Edge FPGA.

%These configurations, both rely on the Kyber KEM, which as demonstrated in work~\cite{} performs faster than the ECDH.
% Additionally as expected, the execution time of the attestation process is notably larger in the less powerful SoC FPGA. However, the overhead added is still acceptable.

\subsection{Security Assessment}

\subsubsection{PQC Algorithms Selection}

We have chosen Falcon~\cite{fouque2018falcon} and Dilithium~\cite{bos2018crystals} as our DSA, and Kyber~\cite{ducas2018crystals} and McEliece~\cite{mceliece1978public} for our KEM. While PQC standards are still in the process of being finalized, these algorithms are considered strong candidates for the upcoming NIST standards. For the hash function (SHA3-512) and the encryption/decryption algorithm (AES-256), we have opted for classical cryptographic methods (with high bit-width), \textcolor{note}{as for our requirements no PQC options are available.} 
% \textcolor{note}{This selection however still maintains a high level of security within our system, as prior research~\cite{bonnetain2019quantum} has demonstrated that these algorithms are effective in countering quantum attacks.}
This selection still maintains a high level of security, as based on research works~\cite{jang2024quantum,roma2021energy}, they are effective in countering quantum attacks.

\subsubsection{Third party attacker}
In case an adversary attempts to load a malicious kernel to the edge node, since the reference values in the attestation server differ, the attestation fails and the malicious kernel isn't programmed to the FPGA. 
% This also implies, in case the adversary tries to insert malicious functionalities in the original code, i.e. Hardware Trojans~\ref{???}. 
% In such case, the checksum $C_3$ calculated in the FPGA will not match the reference, thus the attestation will fail once again.
Furthermore, since the user's bitstream is encrypted, it prevents any reverse engineering attempts by malicious third parties. Additionally, by signing each attestation report $A_i$ before it is shared, we mitigate the risk of man-in-the middle attacks that could alter the report's contents. Lastly, by including a random nonce in every transaction we prevent any replay attacks, in which an adversary would attempt to obtain and re-transmit previous attestation reports (targeting either the attestation server or the blockchain infrastructure).

\subsubsection{Malicious FPGA Operator}
In case an attacker attempts to insert malicious functionalities in the provided infrastructure that could compromise the attestation process, by verifying the attestation service deployed by the operator (phase II, steps \Circled{3}-\Circled{4}), we guarantee the integrity of the edge nodes.
% Furthermore, by using a pre-installed encryption key ($K_{FPGA}$) we also verify the authenticity of the (far) edge device.
\section{Conclusions}

Ensuring the secure configuration of FPGA-based edge nodes in the PQ-era poses new challenges. 
In this work, we presented a solution for ensuring the secure deployment of applications in FPGAs. 
It is built upon remote attestation and combines software and hardware modules to ensure the authenticity of each component. Furthermore, we integrate a blockchain infrastructure to our system, as a trusted storage for any evidence collected. Our evalution over two FPGA families shows that by employing PQC algorithms for the DSA and KEM steps, the execution time overhead is negligible ($2\%$). % with Dilithium and Kyber outperforming other options.
%As future steps, we will explore the integration of Trusted Execution Environments (TEEs), e.g. Intel's SGX, to enhance the security of the software components.
% --------------------------------------------------------------------------

\section*{Acknowledgments}

Authors have been partially supported by the European Union’s Horizon 2020 Research and
Innovation programme PRIVATEER under Grant Agreement No 101096110.

\begin{comment}
This work has been partially funded by the PRIVATEER project.
PRIVATEER has received funding from the Smart Networks and Services
Joint Undertaking (SNS JU) under the European Union’s Horizon Europe
research and innovation programme under Grant Agreement No. 101096110.
Views and opinions expressed are however those of the author(s) only and do
not necessarily reflect those of the European Union or SNS JU. Neither the
European Union nor the granting authority can be held responsible for them.
\end{comment}

% --------------------------------------------------------------------------
\bibliographystyle{ieeetr}
\bibliography{references/references}

\begin{thebibliography}{10}

\bibitem{popovski2019wireless}
P.~Popovski, {\v{C}}.~Stefanovi{\'c}, J.~J. Nielsen, E.~De~Carvalho, M.~Angjelichinoski, K.~F. Trillingsgaard, and A.-S. Bana, ``{Wireless access in ultra-reliable low-latency communication (URLLC)},'' {\em IEEE Transactions on Communications}, vol.~67, no.~8, pp.~5783--5801, 2019.

\bibitem{abedin2018resource}
S.~F. Abedin, M.~G.~R. Alam, S.~A. Kazmi, N.~H. Tran, D.~Niyato, and C.~S. Hong, ``Resource allocation for ultra-reliable and enhanced mobile broadband iot applications in fog network,'' {\em IEEE Transactions on Communications}, vol.~67, no.~1, pp.~489--502, 2018.

\bibitem{samdanis2020road}
K.~Samdanis and T.~Taleb, ``{The road beyond 5G: A vision and insight of the key technologies},'' {\em IEEE Network}, vol.~34, no.~2, pp.~135--141, 2020.

\bibitem{benson2009unraveling}
T.~Benson, A.~Akella, and D.~A. Maltz, ``Unraveling the complexity of network management.,'' in {\em NSDI}, pp.~335--348, 2009.

\bibitem{kreutz2014software}
D.~Kreutz, F.~M. Ramos, P.~E. Verissimo, C.~E. Rothenberg, S.~Azodolmolky, and S.~Uhlig, ``Software-defined networking: A comprehensive survey,'' {\em Proceedings of the IEEE}, vol.~103, no.~1, pp.~14--76, 2014.

\bibitem{hu2015mobile}
Y.~C. Hu, M.~Patel, D.~Sabella, N.~Sprecher, and V.~Young, ``Mobile edge computing—a key technology towards 5g,'' {\em ETSI white paper}, vol.~11, no.~11, pp.~1--16, 2015.

\bibitem{abbasi2020efficient}
M.~Abbasi, A.~Najafi, M.~Rafiee, M.~R. Khosravi, V.~G. Menon, and G.~Muhammad, ``Efficient flow processing in 5g-envisioned sdn-based internet of vehicles using gpus,'' {\em IEEE Transactions on Intelligent Transportation Systems}, vol.~22, no.~8, pp.~5283--5292, 2020.

\bibitem{chamola2020fpga}
V.~Chamola, S.~Patra, N.~Kumar, and M.~Guizani, ``{FPGA for 5G: Re-configurable hardware for next generation communication},'' {\em IEEE Wireless Communications}, vol.~27, no.~3, pp.~140--147, 2020.

\bibitem{barsellotti2022introducing}
L.~Barsellotti, F.~Alhamed, J.~J.~V. Olmos, F.~Paolucci, P.~Castoldi, and F.~Cugini, ``{Introducing data processing units (DPU) at the edge},'' in {\em 2022 International Conference on Computer Communications and Networks (ICCCN)}, pp.~1--6, IEEE, 2022.

\bibitem{la2021denial}
T.~La, K.~Pham, J.~Powell, and D.~Koch, ``Denial-of-service on fpga-based cloud infrastructures—attack and defense,'' {\em IACR Transactions on Cryptographic Hardware and Embedded Systems}, pp.~441--464, 2021.

\bibitem{chakraborty2013hardware}
R.~S. Chakraborty, I.~Saha, A.~Palchaudhuri, and G.~K. Naik, ``Hardware trojan insertion by direct modification of fpga configuration bitstream,'' {\em IEEE Design \& Test}, vol.~30, no.~2, pp.~45--54, 2013.

\bibitem{ender2020unpatchable}
M.~Ender, A.~Moradi, and C.~Paar, ``The unpatchable silicon: a full break of the bitstream encryption of xilinx 7-series $\{$FPGAs$\}$,'' in {\em 29th USENIX Security Symposium (USENIX Security 20)}, pp.~1803--1819, 2020.

\bibitem{bernstein2017post}
D.~J. Bernstein and T.~Lange, ``Post-quantum cryptography,'' {\em Nature}, vol.~549, no.~7671, pp.~188--194, 2017.

\bibitem{zhao2022shef}
M.~Zhao, M.~Gao, and C.~Kozyrakis, ``Shef: Shielded enclaves for cloud fpgas,'' in {\em Proceedings of the 27th ACM International Conference on Architectural Support for Programming Languages and Operating Systems}, pp.~1070--1085, 2022.

\bibitem{oh2021meetgo}
H.~Oh, K.~Nam, S.~Jeon, Y.~Cho, and Y.~Paek, ``Meetgo: A trusted execution environment for remote applications on fpga,'' {\em IEEE Access}, vol.~9, pp.~51313--51324, 2021.

\bibitem{wang2024towards}
Y.~Wang, X.~Chang, H.~Zhu, J.~Wang, Y.~Gong, and L.~Li, ``Towards secure runtime customizable trusted execution environment on fpga-soc,'' {\em IEEE Transactions on Computers}, 2024.

\bibitem{zeitouni2021trusted}
S.~Zeitouni, J.~Vliegen, T.~Frassetto, D.~Koch, A.-R. Sadeghi, and N.~Mentens, ``Trusted configuration in cloud fpgas,'' in {\em 2021 IEEE 29th Annual International Symposium on Field-Programmable Custom Computing Machines (FCCM)}, pp.~233--241, IEEE, 2021.

\bibitem{zhang2024bcae}
S.~Zhang, Z.~Yan, W.~Liang, K.-C. Li, and B.~Di~Martino, ``Bcae: A blockchain-based cross domain authentication scheme for edge computing,'' {\em IEEE Internet of Things Journal}, 2024.

\bibitem{coker2011principles}
G.~Coker, J.~Guttman, P.~Loscocco, A.~Herzog, J.~Millen, B.~O’Hanlon, J.~Ramsdell, A.~Segall, J.~Sheehy, and B.~Sniffen, ``Principles of remote attestation,'' {\em International Journal of Information Security}, vol.~10, pp.~63--81, 2011.

\bibitem{chen2016report}
L.~Chen, L.~Chen, S.~Jordan, Y.-K. Liu, D.~Moody, R.~Peralta, R.~A. Perlner, and D.~Smith-Tone, {\em Report on post-quantum cryptography}, vol.~12.
\newblock US Department of Commerce, National Institute of Standards and Technology~…, 2016.

\bibitem{besu}
``{Welcome | Besu documentation}.''

\bibitem{streit2021choice}
F.-J. Streit, P.~Kr{\"u}ger, A.~Becher, J.~Schlumberger, S.~Wildermann, and J.~Teich, ``Choice--a tunable puf-design for fpgas,'' in {\em 2021 31st International Conference on Field-Programmable Logic and Applications (FPL)}, pp.~38--44, IEEE, 2021.

\bibitem{nejatollahi2019post}
H.~Nejatollahi, N.~Dutt, S.~Ray, F.~Regazzoni, I.~Banerjee, and R.~Cammarota, ``Post-quantum lattice-based cryptography implementations: A survey,'' {\em ACM Computing Surveys (CSUR)}, vol.~51, no.~6, pp.~1--41, 2019.

\bibitem{malina2023deploying}
L.~Malina, P.~Dobias, J.~Hajny, and K.-K.~R. Choo, ``On deploying quantum-resistant cybersecurity in intelligent infrastructures,'' in {\em Proceedings of the 18th International Conference on Availability, Reliability and Security}, pp.~1--10, 2023.

\bibitem{fouque2018falcon}
P.-A. Fouque, J.~Hoffstein, P.~Kirchner, V.~Lyubashevsky, T.~Pornin, T.~Prest, T.~Ricosset, G.~Seiler, W.~Whyte, Z.~Zhang, {\em et~al.}, ``Falcon: Fast-fourier lattice-based compact signatures over ntru,'' {\em Submission to the NIST’s post-quantum cryptography standardization process}, vol.~36, no.~5, pp.~1--75, 2018.

\bibitem{bos2018crystals}
J.~Bos, L.~Ducas, E.~Kiltz, T.~Lepoint, V.~Lyubashevsky, J.~M. Schanck, P.~Schwabe, G.~Seiler, and D.~Stehl{\'e}, ``Crystals-kyber: a cca-secure module-lattice-based kem,'' in {\em 2018 IEEE European Symposium on Security and Privacy (EuroS\&P)}, pp.~353--367, IEEE, 2018.

\bibitem{ducas2018crystals}
L.~Ducas, E.~Kiltz, T.~Lepoint, V.~Lyubashevsky, P.~Schwabe, G.~Seiler, and D.~Stehl{\'e}, ``Crystals-dilithium: A lattice-based digital signature scheme,'' {\em IACR Transactions on Cryptographic Hardware and Embedded Systems}, pp.~238--268, 2018.

\bibitem{mceliece1978public}
R.~J. McEliece, ``A public-key cryptosystem based on algebraic,'' {\em Coding Thv}, vol.~4244, pp.~114--116, 1978.

\bibitem{jang2024quantum}
K.~Jang, S.~Lim, Y.~Oh, H.~Kim, A.~Baksi, S.~Chakraborty, and H.~Seo, ``Quantum implementation and analysis of sha-2 and sha-3,'' {\em Cryptology ePrint Archive}, 2024.

\bibitem{roma2021energy}
C.~A. Roma, C.-E.~A. Tai, and M.~A. Hasan, ``Energy efficiency analysis of post-quantum cryptographic algorithms,'' {\em IEEE Access}, vol.~9, pp.~71295--71317, 2021.

\end{thebibliography}

\end{document}